\documentstyle[12pt,fleqn]{article}

\font\twlgot =eufm10 scaled \magstep1
\font\egtgot =eufm8
\font\sevgot =eufm7

\font\twlmsb =msbm10 scaled \magstep1
\font\egtmsb =msbm8
\font\sevmsb =msbm7

\newfam\gotfam
\def\pgot{\fam\gotfam\twlgot}
\textfont\gotfam\twlgot
\scriptfont\gotfam\egtgot
\scriptscriptfont\gotfam\sevgot
\def\got{\protect\pgot}

\newfam\msbfam
\textfont\msbfam\twlmsb
\scriptfont\msbfam\egtmsb
\scriptscriptfont\msbfam\sevmsb

\def\pBbb{\relax\ifmmode\expandafter\Bb\else\typeout{You cann't use
Bbb in text mode}\fi}
\def\Bb #1{{\fam\msbfam\relax#1}}

\def\op#1{\mathop{{\it\fam0} #1}\limits}

\newcommand{\nm}[1]{\mid {#1}\mid}

\newcommand{\bite}{\begin{itemize}}
\newcommand{\eite}{\end{itemize}}
\newcommand{\benu}{\begin{enumerate}}
\newcommand{\eenu}{\end{enumerate}}
\newcommand{\bde}{\begin{description}}
\newcommand{\ede}{\end{description}}
\newcommand{\bquo}{\begin{quote}}
\newcommand{\equo}{\end{quote}}
\newcommand{\bquot}{\begin{quotation}}
\newcommand{\equot}{\end{quotation}}

\newcommand{\eqref}[1]{(\ref{#1})}
\newcommand{\beq}{\begin{equation}}
\newcommand{\eeq}{\end{equation}}
\newcommand{\ben}{\begin{eqnarray}}
\newcommand{\een}{\end{eqnarray}}
\newcommand{\be}{\begin{eqnarray*}}
\newcommand{\ee}{\end{eqnarray*}}
\newcommand{\bea}{\begin{eqalph}}
\newcommand{\eea}{\end{eqalph}}

\newcommand{\gT}{{\got T}}

\newcommand{\cL}{{\cal L}}

\newcommand{\cE}{{\cal E}}

\newcommand{\cF}{{\cal F}}

\newcommand{\cS}{{\cal S}}

\newcommand{\bL}{{\bf L}}

\newcommand{\al}{\alpha}
\newcommand{\bt}{\beta}
\newcommand{\dl}{\delta}
\newcommand{\la}{\lambda}

\newcommand{\x}{\xi}

\newcommand{\m}{\mu}
\newcommand{\n}{\nu}

\newcommand{\G}{\Gamma}

\newcommand{\ve}{\varepsilon}

\newcommand{\si}{\sigma}

\newcommand{\wt}{\widetilde}

\newcommand{\dr}{\partial}

\newcommand{\ot}{\otimes}
\newcommand{\ap}{\approx}

\newenvironment{eqalph}{\stepcounter{equation}
\setcounter{equationa}{\value{equation}}
\setcounter{equation}{0}

\begin{eqnarray}}{\end{eqnarray}\setcounter{equation}{\value{equationa}}}

\newcommand{\mar}[1]{}

\begin{document}
\hbox{}

{\parindent=0pt

{\large\bf ENERGY-MOMENTUM TENSORS IN GAUGE THEORY} 
\bigskip

{\large \sc  Gennadi
Sardanashvily}
\bigskip

Department of Theoretical Physics, Moscow State
University

sard@grav.phys.msu.su 

http://webcenter.ru/$\sim$sardan/

\bigskip

{\small

{\bf Summary}

In field theory on a
fibre bundle $Y\to X$, an energy-momentum current is associated to a
lift onto $Y$ of a vector field on $X$. Such a lift by no means is
unique and contains a vertical part. It follows that: 
(i) there are a set of different energy-momentum currents;
(ii) the Noether part of an energy-momentum current is not taken away;
(iii) if a Lagrangian is not
gauge-invariant, the energy-momentum fails to be conserved.
\bigskip
\bigskip
}}

In gauge theory, classical fields are represented by sections of a
fibre bundle $Y\to X$, coordinated by $(x^\la,y^i)$. Their
configuration space  
is the first order jet manifold $J^1Y$ of $Y\to X$ coordinated by
$(x^\la,y^i,y^i_\la)$ (where $y^i_\la$ are coordinates of derivatives
of field functions). A Lagrangian on $J^1Y$
is defined as a density
\be
L=\cL(x^\la,y^i,y^i_\la)d^nx, \qquad n=\dim X. 
\ee

By gauge
transformations are meant bundle automorphisms of $Y\to X$. 
To study Lagrangian conservation laws, it suffices to consider
1-parameter groups $G_u$ of gauge transformations. Their infinitesimal
generators are projectable vector fields 
$u=u^\la(x)\dr_\la +u^i(y)\dr_i$
on $Y\to X$. A Lagrangian $L$ is $G_u$-invariant iff
its Lie derivative $\bL_uL$
along $u$ vanishes. The first variational
formula states the canonical decomposition 
\mar{bC30'}\beq
\bL_uL= 
(u^i-y^i_\m u^\m )\cE_id^nx - d_\la \gT_u^\la d^nx, \qquad
d_\la =\dr_\la + y^i_\la\dr_i +y^i_{\la\m}\dr^\m_i, \label{bC30'}
\eeq
where $\cE_i= (\dr_i\cL- d_\la\dr^\la_i\cL)$
is the Euler--Lagrange operator and
\mar{Q30}\beq
\gT^\la_u =(u^\m y^i_\m-u^i)\dr^\la_i\cL -u^\la\cL
\label{Q30}
\eeq
is the current along $u$.
On the shell $\cE_i=0$, the first variational formula (\ref{bC30'})
leads to the weak  
identity 
\be
\bL_uL \ap 
- d_\la[(u^\m y^i_\m -u^i)\dr^\la_i\cL -u^\la\cL]. 
\ee
If the Lie derivative $\bL_uL$ 
vanishes, we obtain the 
weak conservation law 
\be
0\ap -d_\la\gT^\la_u 
\ee
of the current $\gT_u$ (\ref{Q30}).

\noindent
{\bf Remark 1}. 
It may happen that a current $\gT$ (\ref{Q30}) takes the
form
\be
\gT^\la= W^\la +d_\m U^{\m\la}, 
\ee
where the term $W$ vanishes on-shell ($W\ap 0$). Then one says that
$\gT$ reduces to a superpotential $U$. 

\noindent
{\bf Remark 2}. 
Background fields do not live in the dynamic shell $\cE_i=0$ and,
therefore, break Lagrangian
conservation laws as follows. Let us consider the product
$Y_{\rm tot}=Y\op\times Y'$
of the above fibre bundle $Y$ of dynamic fields
and a fibre bundle
$Y'$, coordinated by $(x^\la, y^A)$, whose sections are background
fields. A Lagrangian $L$ is defined
on the total configuration space $J^1Y_{\rm tot}$.
Let 
\mar{l68}\beq
u=u^\la(x)\dr_\la + u^A(x^\mu,y^B)\dr_A + u^i(x^\mu,y^B, y^j)\dr_i
\label{l68} 
\eeq
be a projectable vector field on $Y_{\rm tot}$ which also
projects onto $Y'$ because gauge 
transformations of background fields do not depend on the dynamic ones. 
Substitution of (\ref{l68}) in (\ref{bC30'}) leads to
the first variational formula in the presence of background 
fields
\be
&&\bL_u L= 
(u^A-y^A_\la u^\la)\dr_A\cL + \dr^\la_A\cL d_\la (u^A-y^A_\mu u^\mu)
+\\
&&\qquad (u^i-y^i_\la u^\la)\cE_i
-d_\la[(u^\m y^i_\m -u^i)\dr^\la_i\cL -u^\la\cL]. 
\ee
A total Lagrangian $L$ is usually invariant under gauge
transformations of the product $Y\times Y'$. In this case, we obtain  the 
weak identity
\mar{l70}\beq
0\ap (u^A-y^A_\la u^\la)\cE_A
-d_\la[(u^\m y^i_\m -u^i)\dr^\la_i\cL -u^\la\cL] \label{l70}
\eeq
in the presence of background on the dynamic shell $\cE_i=0$. 

Point out the following properties of currents.

(i) $\gT_{u+u'}=\gT_u +\gT_{u'}$.

(ii) Any projectable vector field $u$ on $Y$ projected onto 
the vector field $\tau=u^\la\dr_\la$ on $X$ is written as the sum
$u=\wt\tau +u_V$ of some lift $\wt\tau$
of $\tau$ onto $Y$ and the vertical vector field 
$u_V=u-\wt\tau$ on $Y$. 

(iii) The current along a vertical vector field $u=u^i\dr_i$ on $Y$ is
the Noether current $\gT^\la_u =-u^i\dr^\la_i\cL$.

(iv) The current $\gT_{\wt\tau}$ along a lift $\wt\tau$ onto $Y$
of a vector field $\tau=\tau^\la\dr_\la$ on $X$ is said to be the
energy-momentum current. 

It follows from the items (i) -- (iv) that any current
can be represented by a sum of an energy-momentum
current and a Noether one.

Different lifts $\wt\tau$ and $\wt\tau'$ onto $Y$ of a vector field
$\tau$ on $X$ lead to distinct energy-momentum currents $\gT_{\wt\tau}$
and $\gT_{\wt\tau'}$, whose difference $\gT_{\wt\tau}-\gT_{\wt\tau'}$
is the Noether current along the vertical vector field
$\wt\tau-\wt\tau'$ on $Y$. The problem is that, in general, there
is no canonical lift onto $Y$ of vector fields on $X$, and one
can not take the Noether part away from an energy-momentum current.

There exists the category of so called natural bundles $T\to X$,
exemplified by tensor bundles, which admit the canonical lift
$\wt\tau$ onto $T$ of any vector  
field $\tau$ on $X$. This is the case of space-time symmetries and
gravitation theory. Such a lift is the infinitesimal generator of a
1-parameter group of general covariant transformations of $T$.
The corresponding energy-momentum current $\gT_{\wt\tau}$ is reduced to
the generalized 
Komar superpotential. Other energy-momentum currents
differ from $\gT_{\wt\tau}$ in the Noether ones, but they
fail to be conserved because almost all gravitation Lagrangians are not
invariant under vertical (non-holonomic) gauge transformations.

Let us focus on field models on non-natural bundles $Y\to X$, i.e.,
they possess internal symmetries. 
Then a vector field on $X$ gives rise to $Y$ by means of a connection
on $Y\to X$. 

A connection on a fibre bundle $Y\to X$ is defined as a section
$\G$ of the affine jet bundle $J^1Y\to Y$, and is represented by the
tangent-valued form
\mar{pr1}\beq
\G=dx^\la\ot(\dr_\la +\G^i_\la\dr_i) \label{pr1}
\eeq
It follows that connections on $Y\to X$ make up an affine space
modelled on the space of soldering forms $\si=\si^i_\la
dx^\la\ot\dr_i$. In particular, the difference
of two connections $\G-\G'$ is a soldering form. Given a
connection $\G$ (\ref{pr1}), a vector field $\tau=\tau^\la\dr_\la$ on
$X$ gives rise to the projectable vector field
\be
\G\tau=\tau^\la(\dr_\la +\G^i_\la\dr_i) 
\ee
on a fibre bundle $Y$. The corresponding current 
\be
\gT_{\G\tau}^\la=\tau^\m T^\la_\m= \tau^\m[(y^i_\m
-\G^i_\m)\dr^\la_i\cL-\dl^\la_\m\cL]
\ee
is called the energy-momentum current, while $T^\la_\m$ is said to be
the energy-momentum tensor with respect to a connection $\G$.

\noindent
{\bf Remark 3}. The difference $\gT_{\G\tau}-\gT_{\G'\tau}$ of energy
momentum currents with respect to different connections $\G$ and $\G'$
is the Noether current along the vertical vector field 
$\G\tau-\G'\tau=
\tau^\la(\G^i_\la-\G'^i_\la)\dr_i.$

\noindent
{\bf Remark 4}. Let $Y\to X$ be a trivial bundle and $\G$ a flat
connection on it. There is a coordinate system on $Y$ such that
$\G^i_\la=0$. Then the energy-momentum tensor with respect to $\G$
reduces to the
familiar canonical energy-momentum tensor. The latter, however, is not
preserved 
under gauge transformations, and it is not defined on a non-trivial bundle.

Let us study energy-momentum conservation laws in gauge theory of
principal connections on a principal bundle $P\to X$ with a structure
Lie group $G$. These connections are sections of the
fibre bundle $C=J^1P/G\to X$, coordinated by $(x^\la,a^q_\la)$,
and are identified with gauge potentials. 
Their configuration space is the jet manifold
$J^1C$ coordinated by $(x^\la,a^q_\la,a^q_{\la\m})$. It admits
the canonical splitting 
\be
a_{\la\m}^r = \frac12(\cF^r_{\la\m} +\cS^r_{\la\m})=
\frac{1}{2}(a_{\la\m}^r + a_{\m\la}^r
  - c_{pq}^r a_\la^p a_\m^q) + \frac{1}{2}
(a_{\la\m}^r - a_{\m\la}^r +
c_{pq}^r a_\la^p a_\m^q). 
\ee

Gauge transformation in gauge theory on a principal bundle $P\to X$ are
automorphism of $P\to X$
which are equivariant under the canonical
action of the structure group $G$ on $P$ on the right.
They induce the automorphisms of the bundle of connections $C$ whose
generators read
\mar{e11}\beq
\x=\x^\la\dr_\la+ (\dr_\m\x^r +
c^r_{pq}a^p_\m\x^q-a^r_\la\dr_\m\x^\la)\dr^\m_r, \label{e11}
\eeq
where $\x^r$ are functions on $X$ which play the role of gauge parameters.

Let $L$ be a 
Lagrangian on $J^1C$. One usually requires of $L$ to
be invariant under vertical gauge transformations with the generators
\be
\x=
(\dr_\la \xi^r+c^r_{qp}a^q_\la\xi^p)\dr^\la_r.
\ee
Hence, $L$ is a function of the strength $\cF$. Then the Noether current
\be
\gT^\la_\x=-(\dr_\m \xi^r+c^r_{qp}a^q_\m\xi^p)\dr^{\la\m}_r\cL 
\ee
is conserved. It reduces to the superpotential
form
\be
\gT^\la_\x =\x^r\cE_r^\la + d_\m U^{\m\la}, \qquad
U^{\m\la}=\x^p\dr^{\la\m}_p\cL.
\ee

Given a principal connection $B$ on $P\to X$, there
exists the lift
\mar{e25}\beq
\wt\tau_B=\tau^\la\dr_\la +[\dr_\m(\tau^\la B^r_\la)+c^r_{qp}
a^q_\m (\tau^\la B^p_\la) - a^r_\la\dr_\m \tau^\la]\dr^\m_r. \label{e25}
\eeq
of a vector field $\tau$ on $X$ onto the bundle of connections $C\to X$. 
It is a generator (\ref{e11}) of gauge
transformations of $C$ with the gauge parameters $\x^r=\tau^\la B^r_\la$.

Discovering the energy-momentum current along the lift (\ref{e25}),
we assume that a Lagrangian $L$ of gauge theory 
depends 
on a background metric on $X$. This metric is a section of
the tensor 
bundle $\op\vee^2 TX$ coordinated by $(x^\la,\si^{\m\n})$. 
Following Remark 2, we define 
$L$ on the total configuration space 
$J^1Y=J^1(C\op\times_X \op\vee^2 TX)$.
Given a vector field $\tau$ on $X$, there exists its canonical lift 
\mar{e26}\beq
\wt\tau_g =\tau^\la\dr_\la + 
(\dr_\n\tau^\al\si^{\n\bt}
+\dr_\n\tau^\bt\si^{\n\al})\dr_{\al\bt} \label{e26}
\eeq
onto the tensor bundle $\op\vee^2TX$. Combining
(\ref{e25}) and (\ref{e26}) gives the lift 
\be
\wt\tau_Y = [\tau_g
- a^r_\la\dr_\m \tau^\la\dr^\m_r] + [\dr_\m(\tau^\la B^r_\la)+c^r_{qp}
a^q_\m (\tau^\la B^p_\la)]\dr^\m_r 
\ee
of a vector field $\tau$ on $X$ onto the
product $Y$. The first term in this expression is a local generator of
general covariant transformations, while the second one is that of
vertical gauge transformations.

Let a total Lagrangian $L$ be
invariant under general covariant transformations and vertical gauge
transformations. 
Then using the formula (\ref{l70}),
we obtain the weak identity  
\mar{b3137}\beq
 0\ap \dr_\la\tau^\m t^\la_\m\sqrt{\nm g} -\tau^\m\{_\m{}^\bt{}_\la\}
t^\la_\bt \sqrt{\nm g} - d_\la\gT^\la_B, \label{b3137}
\eeq
where $t^\la_\m$ is the metric energy-momentum tensor,
$\{_\m{}^\bt{}_\la\}$ are the Christoffel symbols  
of a background metric $g$, and  
\mar{b3147}\beq
\gT^\la_B = [\dr^{\la\nu}_r\cL (\tau^\m a^r_{\m\nu}+\dr_\nu\tau^\m a^r_\m)
-\tau^\la\cL] + 
 [-\dr^{\la\nu}_r\cL (\dr_\nu(\tau^\m B^r_\m) + 
c^r_{qp}a^q_\n (\tau^\m B^p_\m)] \label{b3147}
\eeq
is the energy-momentum current along the vector field (\ref{e25}).
If $L$ is the Yang--Mills Lagrangian, a simple computation brings 
(\ref{b3137}) into the familiar covariant conservation law 
\mar{C112}\beq
\nabla_\la (t^\la_\m\sqrt{\nm g})\ap 0, \label{C112}
\eeq
independent of a connection $B$.
All other energy-momentum conservation laws differ from
(\ref{C112}) in a superpotential term $d_\m d_\la U^{\m\la}$.

If a Lagrangian $L$ is not gauge-invariant, no energy-momentum is conserved.
For instance, let
\be
L =\frac{1}{2k}a^G_{mn}\ve^{\al\la\m} a^m_\al (\cF^n_{\la\m}
-\frac13c^n_{pq}a^p_\la a^q_\m)d^3x
\ee
be the Chern--Simons Lagrangian. It is not gauge-invariant, but 
its Euler--Lagrange operator is so. Then we obtain the 
conservation law 
\be
0\ap -d_\la[\gT^\la_B + \frac1ka^G_{mn}\ve^{\al\la\m}\dr_\al(\tau^\nu
B^m_\nu) a^n_\m], 
\ee
where $\gT_B$ is the energy-momentum current (\ref{b3147}) along the
vector field $\wt\tau_B$.
Thus, the energy-momentum current of the Chern--Simons model
is not conserved, but there exists another conserved quantity.

Another interesting example is a
Lagrangian in the generating
functional of quantum gauge theory. It is not gauge-invariant, but is
BRST-invariant. The corresponding energy-momentum current is conserved,
but it contains ghost fields.

\bigskip
\bigskip

{\parindent=0pt

{\bf References}

G.Sardanashvily, Energy-momentum conservation laws in
gauge theory with broken invariance, {\it E-print arXiv}: {\bf hep-th/0203275}

L.Mangiarotti and G.Sardanashvily, {\it Connections in
Classical and Quantum Field Theory} (World Scientific, Singapore, 2000).

G.Sardanashvily, Stress-energy-momentum tensors in 
constraint field theories, {\it J. Math. Phys.} {\bf 38} (1997) 847. 

G.Sardanashvily, Stress-energy-momentum conservation law in 
gauge gravitation theory, {\it Class. Quant. Grav.} {\bf 14}  (1997) 1371.

}

\end{document}